\documentclass [12pt]{article}
\usepackage{amsmath}
\usepackage{amssymb}
\usepackage {graphicx}

\begin{document}

\title{New Scenario to Chaos Transition in the Mappings with
Discontinuities}

\author{ S.V. Naydenov$^1$, A.V. Tur$^2$,
A.V. Yanovsky$^3$, V.V. Yanovsky$^1$ \\ $^1${\it Institute for
Single Crystals,} \\ {\it 60 Lenin avenue, 61001 Kharkov,
Ukraine},
\\ $^2${\it Center d'Etude Spatiale Des Rayonnements,
C.N.R.S.-U.P.S.,} \\ {\it 9 avenue du Colonel Roche, 31028
Toulouse, CEDEX 4, France} \\ $^3${\it Institute for
Low-Temperatures Physics and Engineering,} \\ {\it 47 Lenin ave.,
61164 Kharkov, Ukraine} }

\date{\empty}

\maketitle

\begin{abstract}
We consider a many-parametric piecewise mapping with
discontinuity. That is a one dimensional model of singular dynamic
system. The stability boundary are calculated analytically and
numerically. New typical features of stable cycle structures and
scenario to chaos transition provoked by discontinuity are found.



\end{abstract}

\section{Introduction}\label{S1}
It is well known that the dynamical chaos is taken as a basis for
the description of complex behavior in different physical systems
(see for example \cite{Shuster},\cite{ChaosBook} and references
therein). The transition from regular to chaotic dynamics is
typical to dynamical systems with a few number of degrees of a
freedom \cite{LichtLiber}. Even simple dynamic systems can have a
complex non-regular behavior \cite{May}. The phenomenon of
dynamical chaos is specific not only for classical physics but
also for quantum systems \cite{Gutzwiller},\cite{Stock}.

The study of ``turbulence'' mechanisms for waves and oscillations
of different kind processes is one of the more important point for
dynamical chaos. To explain the turbulence origins some mechanisms
has been proposed: the Landau--Hopf scenario \cite{Landau},
\cite{Hopf*}; scenario of Ruelle--Takkens \cite{RuelTakk} with a
strange attractor \cite{Lorenz}; crisis with quasi periodicity
losses \cite{QuasiPeriod}; intermittence transitions
\cite{BerPomoVid}; etc. The instability which appears due to the
non-linearity and the exponential sensitivity of phase trajectory
to initial conditions is crucial for all these cases. The
transition to chaos scenario can be a rough one, but the stability
loss is going continuously. As a rule the Lyapunov index depends
analytically of the control parameter in the bifurcation point.

The more simple and natural models of dynamical chaos are mappings
which can be considered as a finite dimensional approximation of
real dynamical systems. There are a lot of typical scenario to
chaos transition in mappings corresponding to different turbulence
mechanisms. For example: period doubling cascade with Fiegenbaum
universality \cite{Double},\cite{Feigen}, intermittency
\cite{Intermittency}, strange attractor or repeller appearance
\cite{StrAttr},\cite{StrRepp}. At the same time in the same
mapping, but in different control parameter domains, one can
observe different kinds of chaos transition. One dimensional and
two dimensional one-parametrical mappings are widely used, such
as: Bernoulli shift \cite{Billingsley} and ``saw'' \cite{Renyi},
logistic mapping \cite{Logos}, Ulam's ``tent'' \cite{Tent}, twist
transformation \cite{Rotation}, ``backer'' transformation
\cite{Ott}, Henon's transformation \cite{Henon}, Arnold's ``cat''
\cite{CatArn} and ``standard'' Chirikov mapping \cite{Univers}
etc. The one-dimensional continuous mappings and dynamical systems
with one control parameter are studied well enough (see for ex.
\cite{Shark}--\cite{CollEckmLandf}). Bifurcations in bi- and many
parametric families are less studied than others. The instability
and the chaos transition are generally connected with a smooth
non-linearity, a non-monotonic behavior and with a mapping
ambiguity. Each of this factors is necessary, but not sufficient
conditions for the loss of a stability. The dynamics of mappings
with singularities has its special features. Such mappings appear
during the description of impact
oscillator~\cite{Impact1}--\cite{Impact2}, during the modeling of
noise influence on information
capacity~\cite{Bollt}--\cite{ZycsBollt}, during the research of
different dynamic systems with discontinuous characteristics.

The role of such particularities as a break value in
one-dimensional discontinuous mapping of interval has not been
studied in details, especially for a case, when the break value is
less than the interval length (non-overall break). At the same
time, the topological discontinuity and stretch non-linearity
together lead to some new particularity of the dynamical behavior.
In this paper we are considering such features for the simple
piecewise linear mapping case with discontinuity.

The most important new feature of mappings with discontinuity as
it will be shown latter is a new bifurcation of cycle deleting,
which we will call ``put-out'' bifurcation. This bifurcation is
different from other ones in smooth dynamical systems by the
jump-like variation of multiplicator (mapping Jacobian). As a
result of such a bifurcation the stable limit cycle can simply
disappear. Afterwards, the system can get immediately in the
global chaos state. Roughly speaking the ``put-out'' bifurcation
appears as a result of coincidence of one cycle point with break
coordinates. The mapping which we consider is two-parametric and
that is why it has a rich enough dynamic, both regular and
chaotic.

\section{Mapping with a discontinuity}\label{S2}
Let us consider the family of piecewise linear mappings for the
interval $I=[0,1]$:
\begin{equation}\label{1}
x\rightarrow f(x)=\sum _{\sigma }f_\sigma (x)\chi _{\sigma }(x)
;\;f_\sigma (x)=k _\sigma x+b_\sigma \,,
\end{equation}
where $\chi _\sigma$ -- is indicator of the natural partition of
phase space $I=\bigcup _\sigma I_\sigma$ for function $f$ monotony
(linearity) intervals. We will restrict our study to bimodal (with
two critical points) mappings with one break only. For our mapping
we have:
\begin{equation}\label{2}
I_\sigma :\; I_1=\left[0,a\right];\;I_2=\left[a,1/2\right];\;
I_3=\left[1/2,1-a \right];\;I_4=\left[1-a,1 \right] \,,
\end{equation}
where $x=a$ and $x=1-a$ are critical points; $x=c=1/2$ -- is a
break coordinate. Mapping parameters
\begin{equation}\label{3}
\left\{\begin{array}{l}
  k_1=k_4=\mu _1=\frac{A}{a} \\
  k_2= k_3=\mu _2=\frac{2(E-A)}{1-2a}
\end{array}\right. ;\;
\left\{ \begin{array}{l}
  b_1=0;\; b_4=1-\frac{A}{a} \\
  b_2=\frac{A-2aE}{1-2a};\; b_3=1+\frac{A-2(1-a)E}{1-2a}
\end{array}\right.
\end{equation}
include a critical value $A$, $0\leq A=f(a)\leq 1$ and a break
height $E={1+\varepsilon }/{2}$, where $1/2\leq E\leq 1$ (with a
gap $0\leq \varepsilon \leq 1$). The plot of the mapping is shown
on fig. 1.

The mapping (\ref{1})--(\ref{3}) with symmetry (involution) $G$ is
reversible mapping
\begin{equation}\label{4}
f\circ {G}={G}\circ f \Leftrightarrow f(x)=1-f(1-x) ;\;
G(x)=\tilde{x}\stackrel{def}{\equiv}1-x \,,
\end{equation}
where ${G}\circ {G}={id}$; $\circ$ is mappings composition, $id$
-- identical transformation. The reversible systems are important
from physical point of view and a lot of papers treats this
subject \cite{ArnSevr}--\cite{Qwisp}, etc.

The mapping family (\ref{1}) is three parametric $f=f_\Omega$, but
further it will be enough to consider a two parametric case with
$a=const$. It seems convenient to choose as independent
coordinates on $\Omega _a$ parameter surface a set of $(A,E)$. A
necessary but not sufficient condition for chaotic behavior is a
stretching condition so as to at least one of the multiplicators
modules $\left|\mu _{1,2}\right|$ does exceed one. Let us choose
$A> a \Leftrightarrow \mu _1>1$. Because of reversibility we can
consider that $0<a\leq 1/2$. The mapping becomes non-monotonous
and non- one-to-one at $\mu _2 <0$, i.e. lower ($E<A$) than the
diagonal $\Delta$ ($A=E$) of the parametrical space.

The mapping iteration dynamic as well as its bifurcation must be
considered in total space $Z=I\times \Omega$. For one dimensional
system we can consider only the section of the space $Z$
(stability diagram) and bifurcation diagrams on hyper planes
$I\times {D}$ ($D$ -- the value of control parameter in one
parametric family of mappings given by equation $\Phi (A,E;a)=0$).
This last case concerns also bifurcation diagrams in coordinates
$(x,A)$, $E=const$; $(x,E)$, $A=const$ and $(x,\Delta )$, where
$\Delta =A\equiv E$ is a parameter along the diagonal of surface
$\Omega _a$. Periodical trajectories (cycles) play a key role for
the dynamics of one dimensional mappings. Their structure and
stability determine the regular and the chaotic dynamic of a
system.

\section{Cycles structure}\label{S3}
Coordinates of arbitrary cycle $C=\left\{x_i\right\}_i$,
$i=0,\ldots ,p-1$ are defined by equations $f^p(x_i)= x_i$, with
conditions $f^k(x_i)\neq x_i$, $k=1,\ldots ,p-1$. All cycle points
are different $x_i\neq x_j$, $i\neq j$ due to the uniqueness of
the phase trajectory. Point cycle numbering can be changed
cyclically.

Each cycle $C$ is given by a set of characteristic parameters such
as a period $p\geq 2$, a kind $(n,k)$ ($p=n+k$) and itinerary $W$
(topological type). The kind of cycle is determined as a number of
points that belong to monotony segments in neighborhood of
discontinuity, i.e. to segments $I_2$ and $I_3$ of nature
partition. The cycle itinerary $W(C)=\left\{ \sigma [x_i]
\right\}_{i}$ is defined by the cyclic sequence of its point
addresses in nature partition $I$, where $\sigma [x]=\sum_{\sigma
}\sigma \, \chi _\sigma (x)$ is an address which has one of four
possible values $\sigma =1,2,3,4$, for example. In addition, the
points itinerary of this cycle $W(x_i)=\left\{ \sigma [f^k(x_i)]
\right\}_{k}$ is also important. They are formed by different
permutations of the given itinerary $W(x_0)$ with a fixed initial
point $x_0$. It is necessary to differ cycle itinerary from point
one.

The cycle stability is determined by multiplicators $\mu (x)=
df(x)/dx$ in each of its points. According to chain rule of
differentiation we have:
\begin{equation}\label{5}
\mu (C)=\mu ^{(p,k)}(x_i)=\prod _{k=0}^{p-1} df (f^k(x_i))/dx =
\prod  _{k=0}^{p-1} \mu _{\sigma [x_k]}= \mu _1^{p-k}\mu _2^k \,,
\end{equation}
where the cycle multiplicator does not depend of chosen point
$x_i$. It is convenient to introduce the multiplicator module
$m=|\mu |$. For stable $C^s$ cycle $0\leq m(C^s)<1$, for unstable
$C^u$ cycle $m(C^u)>1$. On the stability limit $m(C)=1$ and
additional analysis of stability is required.

Let us describe in details the cycle structure of our mapping. In
accordance with the symmetry (\ref{4}) all cycles can be dived
into symmetric and non symmetric ones. For non symmetric cycles
($\alpha$-type) we have $C_\alpha\neq {G}(C_\alpha )=
\tilde{C}_\alpha $. The symmetric cycle ($\beta$-type) coincides
with its dual one $C_\beta = {G}(C_\beta )= \tilde{C} _\beta $.
Because of reversibility equation $\tilde{\tilde{C}}=C$ is valid
for any cycle. The non symmetric cycles of reversible mappings
appear as dual pairs only $C_\alpha$ and $\tilde{C}_\alpha $. Dual
cycles multiplicators coincide $\mu (C)=\mu (\tilde{C})$. This is
why it is enough to show on stable cycle diagram only one of dual
cycle. On bifurcation diagram dual cycles are symmetric with
respect to $x=1/2$ right line.

Cycles of the same period ${p}$ constitute ${p+1}$ non-overlapping
(non crossing) classes in accordance with possible values
$k=0,\ldots ,p$. For the family of piecewise mappings this
partition is simple: each class consists of only one specimen
(with the accuracy for reversibility $C\rightarrow \tilde{C}$).
The kind $k=1$ or $k=2$ corresponds to non symmetric or symmetric
cycle correspondingly. The piecewise mappings cycles including the
break ones have some particularities. The topological type $W(C)$
determines all the cycle C without ambiguity. The reason is that
the linear equation system for definition of cycle points may have
only one unique solution. In addition, for the same reason there
are no two different cycles for the same period $p_1=p_2
\Rightarrow C_1=C_2$. Cycles have following isomorphic
descriptions
\begin{equation}\label{6}
C(p;x_0)\leftrightarrow W(C)\leftrightarrow \left\{\mu (C);\,
k\right\}\,.
\end{equation}

All stratified cycle structure may be shown as
\[C=C_{asymm} \oplus C_{symm};\; C=C^{stab}\oplus C^{nonstab}\;;\]
\[C^{stab}= C_{asymm}(k=1)\oplus C_{symm}(k=2) \;;\]
\begin{equation}\label{7}
C=\bigoplus _{p=2}^{\infty} C(p)= \bigoplus
_{p=2}^{\infty}\bigoplus _{k=0}^p C(p,k)=\bigoplus _{W}C(W) \,,
\end{equation}
where $\oplus$ means the direct sum of subspaces. It is shown
schematically on fig.~2. Mapping is bimodal, that is why stable
cycles higher then $k \geq 3$ kind do not exist. Non-symmetrical
stable cycles of kind $k=1$ can be of even and odd period $p\geq
k$. Symmetric stable cycles have kind $k=2$ and even period only.
Stable cycles of zero $k=0$ do not exist in such a system. Stable
cycles of different itinerary can not coexist. In the contrary the
coexistence of different type of non stable cycles is not
forbidden. Outside of stability area all non symmetrical stable
cycles are situated below the diagonal $A=E$. Symmetrical stable
cycles of the same type are filling two one-link domains on
parameter surface separated by stability area. This is due to the
monotony of the mapping with $A>E$. All the cited regularities are
proved directly or verified by numerical experiments fig.~3,
fig.~4.

\section{Put-out bifurcation}\label{S4}
Reconstruction of cycle structure is determined by ``put-out''
bifurcation which is an important point for dynamic of system with
break. The ``put-out'' bifurcation means the reconstruction
leading to sudden disappearance or appearance of cycles. In a
general case we will define the ``put-out'' bifurcation as a
bifurcation accompanied by jump like changing of mapping
multiplicator (Jacobian) in any point of cycle. It means
\begin{equation}\label{8}
C_1 \rightarrow C_2 \;;\quad \left| \mu (C_1) \right| \neq \left|
\mu (C_2) \right| \,.
\end{equation}
The breaking makes the essential difference between the
``put-out'' bifurcation and usual continuous bifurcation cycles
including ``fork'' bifurcation with doubling of the period and
disappearance of fixed point (cycle) for the intermittency on the
stability threshold.

Let us consider the topological reasons of ``put-out''
bifurcation. ``Special'' trajectories with initial points in the
critical value of mapping or discontinuity are very important for
mapping with singularities. With some values of parameter they
become periodical (cycles). Special cycles give Markovian
partitions for space phase. The dynamic of all others cycles must
be coherent and is controlled by these partitions. One of the
point of special cycle may coincide with any point of usual one
when control parameter are changing. In this case the jump-like
changing of the multiplicator takes place and the ``put-out''
bifurcation appears. Special cycles coordinates of different type
are defined, for our case, by finite number of linear equation.
Special cycles are forming lines of co-dimension one on the plane
of parameters. These lines represent limits of ``put-out''
bifurcation.

The conditions of ``putting-out'' correspond to the cyclicity of a
special trajectory with itinerary $W\left(f^n(\hat{x})\right)$ of
all its points $f^n(\hat{x})$ following. The points correspond to
itinerary $W$ of the cycle of period $p$ under consideration
\begin{equation}\label{9}
f^{p-n}\left(\left. f^n(\hat{x})\,\right|
W(f^n(\hat{x}))\right)=\hat{x} ;\; f^p(\hat{x})=\hat{x} ;\;
\hat{x}=\{a ;\, c \} \,,
\end{equation}
where by $a$ and $c$ the coordinates of critical points and the
points of discontinuity are represented.

The borders of ``put-out'' bifurcation for 2-parametric
($a=const$) family (\ref{1}) should be put down as:
\begin{equation}\label{10}
\left\{
 \begin{array}{l}
  f^{p-n}\left(\left. f^n(a)  \:\right| W(f^n(a))\right)=a \,; \\
  f^{p-n}\left(\left. f^n(1-a)\: \right| W(f^n(1-a))\right)=1-a \,; \\
  f^{p-n}\left(\left. f^n\left(1/2 \mp 0\right) \:
  \right| W\left(f^n\left(1/2 \mp 0\right)\right) \right)
  = \left(1/2 \mp 0\right)\;.
\end{array}\right.
\end{equation}
The itineraries for different points of the same cycle are
different, $W(f^{k_1}(x^*))\neq W(f^{k_2}(x^*))$. Each composition
$f^p$ has $N_p$ one-to-one branches. So all the $N_p=4p$ of the
conditions (\ref{10}) should be regarded as independent. For
stable cycle s or for symmetric cycles the dual conditions,
differing by the change of form (\ref{10}), coincide. Moreover,
for stable cycles all the conditions of every type are equivalent,
so the following independent conditions are left:
\begin{equation}\label{11}
f^{p}\left(\left.a\:\right| W(a)\right)=a ;\;
f^{p}\left(\left.1/2\:\right|W\left(1/2\right)\right)=1/2 \,.
\end{equation}
As an example, let us cite the borders of ``putting-out'' for
$24$, $2431$ and $2441$-cycles (the itineraries are mentioned).
The first of the conditions (\ref{11}) gives $A=A_{24}=1-a$;
$A=A_{2431}=A_{2441}=\frac 12\left[1+\sqrt{1-4a^2}\right]
>A_{24}$; and the second one gives
$E=E_{24}(A)=E_{2431}(A)=1-\frac{a}{2A}$ and
$E=E_{2441}(A)=1-\left(a^2/A^2\right)\left(1-a/\left(2A\right)\right)>1/2$.
The equations $A=A(a)$ ($a=const$) and $E=E(A)$ obtained define
the lines of ``putting-out''. In the general case they are
straight segments ($A=const$) and algebraic curves $E-E_W(A)=0$
(of $p-1$ order for a non-symmetric cycle with period $p$).

Markovian partitions of hyperbolic mappings are everywhere dense
\cite{Sin(Mark)}, \cite{Bowen}. The analogous thing is true for
the mappings (\ref{10}), corresponding to the ``putting-out'' of
unstable cycles of arbitrary high order $p=2,3,\ldots ,\infty $.
Let us note that this lets one calculate the (statistic) invariant
measure of mapping almost everywhere in chaotic region. The
rigorous solution is obtained at the parameters corresponding to
every intersection point of curvilinear coordinate lines of the
bifurcational web of the borders of ``putting-out''. Whereas the
coordinate families mentioned are everywhere dense on the
parametric plane.

``Put-out'' bifurcation is a characteristic feature of not only
one-dimensional systems of the type (\ref{1}), but also of
multi-dimensional and multi-parametric nonlinear systems with and
without breaks. In discontinuous systems, that, according to
Baire, make up a set of the first category (see, for ex.,
\cite{Oxtoby}) in the space of arbitrary dynamic systems -- this
is a typical feature. In continuous systems, the presence of
critical points (singularities) is required. The traces of
``putting-out'' can be found in every discontinuous mapping. The
new mechanisms of transition to chaos, different from those known
before, will be connected with ``putting-out''. The necessary
condition of chaotic behavior is the loss of stability. At that,
in continuous systems, cycles lose stability in a continuous way.
In discontinuous systems, the mechanism of spontaneous loss of
system stability (in large), at ``putting-out'' of a stable cycle
with non-unit module of the multiplier, appears. Such a stable
cycle can be situated far from its border of stability, $\left|\mu
\right| < 1$.

\section{Stability}
\label{S5} Conditions for cycle stability $\mu _1^{p-k}\left|\mu
_2^k\right| <1$ in variables $0< A\leq 1$ and $0<E\leq 1$ ($a <
1/2$ fixed) has the form
\begin{equation}\label{12}
\left( \frac{A}{a}\right)^{p-k}\left| \frac{2(A-E)}{1-2a}\right|^k
<1 \,.
\end{equation}
It corresponds with stability boundaries of cycle $(p,k)$
($p$-period, $k$-kind given by following equations
\begin{equation}\label{13}
E={E_{\pm}}^{p,k}(A)=A\left\{1\pm\left[a^{\frac{p-k}{k}}
\left(1/2-a\right)\right] A^{-\frac{p}{k}}\right\} \,,
\end{equation}
for $1 \geq E > A\geq 1/2$ or $1/2 \leq E < A \leq 1$. Let us
note, that stability boundaries for symmetric stable cycles may
coincide with a part of boundary with the ``put-out'' of
non-symmetric and non-stable cycles (see fig.~3).

With fixed $p$ and $k$ stability boundaries (\ref{13}) are
algebraic curves of $p+k$ order. They limit stability zone in the
square $(1/2\leq A\leq 1;\, 1/2\leq E)$. The curve $E=E_+(A)$ is
convex (down), $E=E_-(A)$ is concave (from above). On the diagonal
$A=E$ of surface $(A,E)$ every kind of unstable cycle is "super"
stable
\begin{equation}\label{14}
A=E \quad \Rightarrow \quad m(C)=0 \,.
\end{equation}

The surface of limit admissible zone of stability decreases with
the increasing of period ${p}$ i.e. $S(p_1)>S(p_2)$ with
$p_1<p_2$. (The real zone of stability comparing to the limit one
is less important due to cycles ``put-out''). It follows from the
monotonically decreasing dependence $E_{\pm}=E_{\pm}(A;\,p,k)$
from ${p}$ (derivative $\partial E_{\pm}/\partial p <0$) and is in
accordance with the fact that with the increasing of $k$ -kind
fixed stable period cycles their stability decreases. Numerical
experiments still confirm this conclusion when taking into account
``put-out'' cycles. The largest area on the stable cycles diagram
is possessed by the most stable stable cycles of $23$ type, then
stable cycles $24$ and $2431$ and so on.

\section{Chaotisation mechanisms}\label{S6}
The chaotisation mechanism depends on the way of system
stability loss. In continuous system this transition is
continuous. During it, the last stable cycle's left multiplicator
module passes over the unit value. In systems with breaks, as
mentioned above, a cycle can be entirely ``put out`` up to the
loss of stability. (Non-symmetric cycle can be ``put out'' only
preserving the system stability, i.e. by appearing of a new and
already symmetric stable cycle. The cascade of period doubling
(see below) corresponds to it.) ``Putting out'' a still stable
cycle can result in stability loss of the whole system. This is
the \textit{spontaneous chaotisation} mechanism.

A typical feature of continuous one-dimensional mappings is the
cascade of period doubling. In it, at the same time with the
smooth loss of stability by some cycle, another stable cycle with
doubled period arises. Cycle multiplier is changing continuously.
The cascade of doubling is a part of Sharkovsky series
\cite{Shark} of constant mapping for cycles ordering by period. In
mappings with breaks the cascade of doubling is also possible. But
it is caused not by constant stability loss in one of the cycles,
but by its ``putting out'' and replacing it by other stable cycle.
There is a jump between the multiplier of the initial cycle and
that of the cycle with double period. Bifurcation of such type is
called \textit{``putting out'' with doubling}.

The rebuilding during which a stable cycle of a certain type is
replaced by a cycle of some other type but the same period is
forbidden in continuous systems, because this should be
accompanied with a jump of cycle multiplier. In discontinuous
systems such type of stable cycles bifurcation is allowed. It is
called \textit{``putting out'' with period preserving.} Let us
point out that the stable cycle stability is preserved in this
case.

``Putting out'' of unstable cycles with preserving or changing of
the period is possible either in continuous or discontinuous
systems. The permissibility of isolated cycles multiplier jump is
guaranteed by everywhere dense in fill of the chaotic system phase
space with other periodic trajectories. (The averaged multiplier
of the mapping near such bifurcation point is continuous.)
``Putting out'' of the unstable cycles influences the
stochasticity procedure. The density of invariant distribution,
metric entropy and so on are measured by the jump.

The three types of bifurcation mentioned belong to mapping stable
cycles. The route of every stable cycle can be represented as
$W(C^s_\alpha )= suU$ and $W(C^s_\beta ) =
suU\,\tilde{s}\tilde{u}\tilde{U}$, where $s=2,3$ -- the address of
cycle points on the components adjacent to the break; $U$ is a
sequence of other $u=4,1$ -- addresses; $\tilde {s}=3,2$ and
$\tilde {u}=4,1$ are dual addresses. Such representation is
imposed by the exclusion of $k \ge 3$-type stable cycles in the
structure of the cycles (\ref{7}). The exclusion of $s\tilde
{u}$-type path elements follows from the topological dynamics of
the cycles. With the use of these symbols the types of stable
cycles ``putting out'' can be represented as:
\[ 1)\; C^s_{\beta }\leftrightarrow C^u_{\alpha };\;
 m(C^s_\beta )<1<m(C^u_\alpha)\,;\]
\[ 2)\; C^s_\alpha \leftrightarrow C^s_\beta ;\; W( C^s_\beta )=
W(C^s_\alpha )W( \tilde{C^s_\alpha }) \,;\]
\[ m(C^s_\beta )=m^2(C^s_\alpha )< m(C^s_\alpha )<1 ;\;
p_\beta =2p_\alpha \,;\]
\[ 3)\; C^s_\beta \leftrightarrow C^s_\alpha ;\: W(C^s_\beta )= suU
\, \tilde{s}\tilde{u}\tilde{U} \,; \]
\begin{equation}\label{15}
W(C^s_\alpha )= suU \, u\tilde{u}\tilde{U} ;\: m(C^s_\beta )<
m(C^s_\alpha )<1 ;\: p_\beta =p_\alpha \,.
\end{equation}
This is schemed on fig.~5. At spontaneous chaotisation the
symmetric stable cycle disappears. At the same time, far from it
(in the phase space) an asymmetric unstable cycle of the same
period and corresponding route ($\tilde {s}$ replaced by $\tilde
{u}$) is situated. So spontaneous chaotisation corresponds to
global bifurcation of discontinuous mapping stable cycles
``putting out''.

The explicit nature of the types of ``putting out'' under
consideration can be observed on the diagram of mapping stability
(\ref{1}). It is numerically obtained at $a=0.4$ and shown in
fig.~3 and fig.~4. The repetition of bifurcations with stable
cycle period doubling and preserving leads to cascades of
doubling. Every cascade is generated by an ``irreducible'' stable
cycle. Together they combine into a ladder of cycles. Irreducible
are cycles of the type: $24;\, 244;\, 24...4$ (they make the main
section of the ladder); $244144;\, 24414;\, 244141$ and others
(they fill the windows between the cascades of the main series).
The regions of stable cycles are limited by the lines of stability
from above and below. The ladder of stable cycles is of fractal
character. Spontaneous chaotisation takes place at transition from
one of its steps (from the region of stable symmetric cycle) into
the chaotic region (beyond the limits of the next step of doubling
cascade). For example, the transition from $23$-drains zone to the
region above the step that includes zones of $24$- and $2431$-
stable cycles. This transition takes place at
$A=A_{23}=A_{24}=1-a$ and $(2-3a)/2(1-a)<E<(3-4a)/2$ or
$(1-2a)/(1-a)<\varepsilon <2(1-2a)$. For the cycles of a higher
period, similar transitions are possible not only above, but also
below the diagonal $A=E$.

\section{Transitions to chaos}\label{S7}
Let us briefly discuss the features of the dynamics under
consideration. The transition to chaos in a singular system can
take place with different scenarios. They correspond to different
one-parametric deformations in the family of mappings. On the
parametric plane, such deformations define the motion along some
curves. The change of modes corresponds to the intersection of the
``dynamic state'' line with bifurcational curves and the curves of
stability. The way of transition to chaos depends on the three
elementary processes, connected with the change of stability. This
is an ordinary stability loss of the cycle and the ``putting out''
of a stable cycle. ``Putting out'' of a stable cycle can lead to
stability loss for the system in general or doubling of stable
cycle period (and a step-wise decrease of its stability). The
doubling is connected with the break gap variations at the fixed
amplitude is $1/2<A<1$, and the spontaneous chaotisation is
connected with the amplitude variations of the mapping with a gap
$\varepsilon \neq \{0,\, 1\}$.

In continuous systems, chaos is caused by their non-linearity and
stretching (in one-dimensional mappings). In singular systems the
gap iterations dynamics is essential. Filling the phase space, the
gaps can preserve or ruin the regularity of the main phase
trajectories or make them ``double''. So, singularity shows itself
as strong ``non-linearity''. Its distinguishing feature is the
uneven character of the processes that take place when the main
parameters of the trajectories, for instance, the multipliers of
the cycles break in the points of bifurcation. The nonlinearity
and ``putting-out'' are closely connected, because they both are
absent in a linear system. ``Putting-out'' of unstable cycles is
typical for discontinuous systems and can take place in continuous
ones (trajectories thinning). In this case, the singularity
directly influences the internal properties of chaos.

Among chaotisation scenarios for discontinuous mapping one can
point out the following transitions: symmetric stable cycle -
(spontaneous) chaos; non-symmetric stable cycle - (ordinary)
chaos; the cascade of doubling by ``putting-out'' - chaos; stable
cycle - chaos - regularity windows (doubling cascades) - chaos
(``putting-out alternation''). All of them can be easily
identified in the cycles diagram fig.~4 -- fig.~5. The main role
in the chaos development is played by the stable cycles
``putting-out'' caused by the singularity of the system.

\section{Summary}\label{S8}
So, on the example of a piecewise-linear mapping with a break the
typical features of regular and chaotic dynamics of singular
dynamic systems are determined. The cycles ``putting-out''
bifurcation for continuous and discontinuous mappings is obtained.
The ``putting-out'' conditions are obtained and their connection
with Markovian phase space partitions, generated by the system
properties (critical points and breaks) is stated. For
discontinuous mappings, three types of stable cycles
``putting-out'' are distinguished - with the stability loss at
``putting-out'' of the whole stable cycle (spontaneous
chaotisation); preserving stability and doubling or preserving of
a cycle's stable period at the change of symmetry character. The
first case corresponds to the new chaotisation mechanism of
singular dynamic systems. The second one corresponds to the
development of period doubling cascades, caused not by the
stability loss of the cycles, but by their ``putting-out'' from
the full structure of mapping cycles. The cycle multipliers, and
so their parameters, defining the chaoticity of the system
(Lyapunov index etc.), in all other types of ``putting-out'' are
measured by a step.

The research of ``putting-out'' bifurcation and the spontaneous
mechanism of stability loss will help us find out fundamental
regularities of the deterministic chaos is arbitrary singular
dynamic systems.

\newpage

\begin{figure}
\begin{center}
\includegraphics*[scale=0.8]{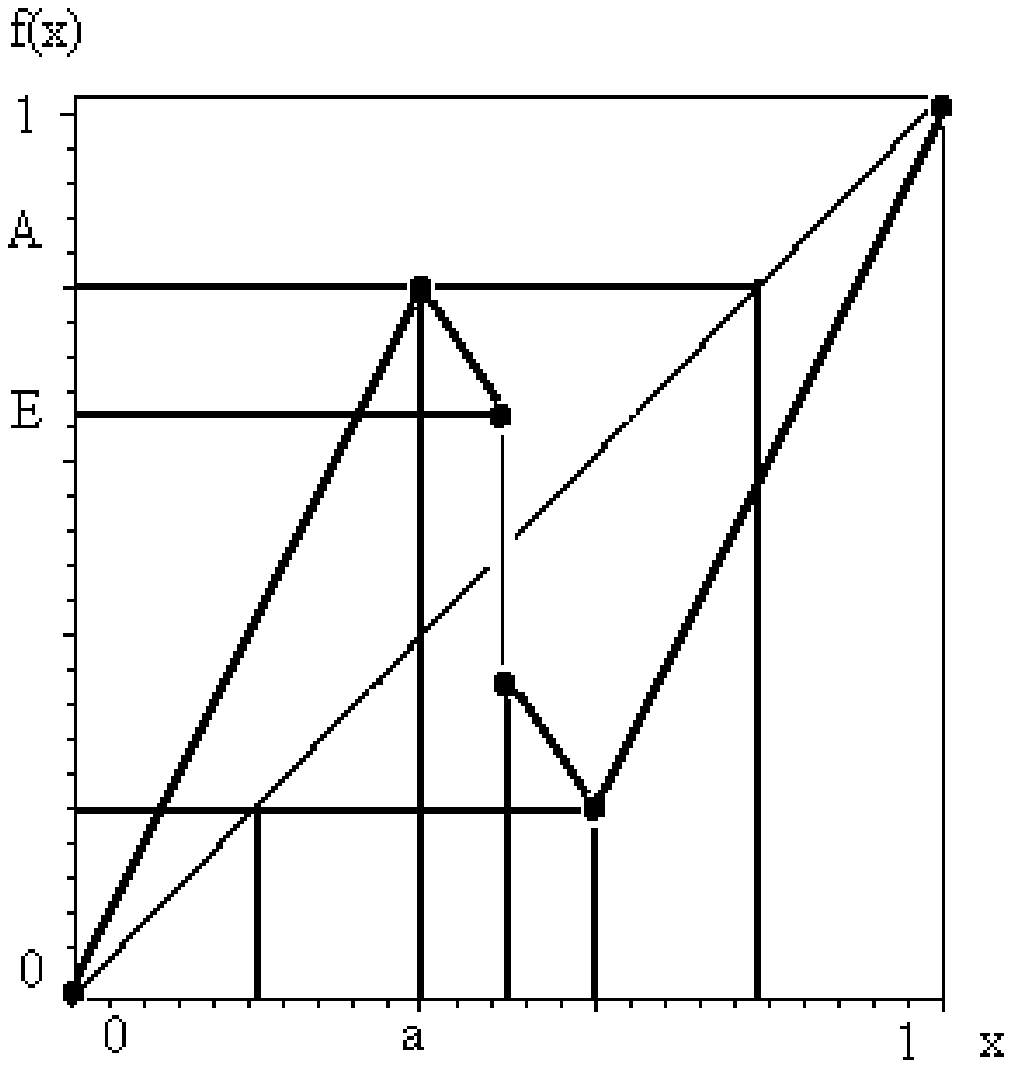}
\end{center}
\caption{Piecewise-linear mapping with a break.}\label{fig1}
\end{figure}

\begin{figure}
\begin{center}
\includegraphics*[scale=0.7]{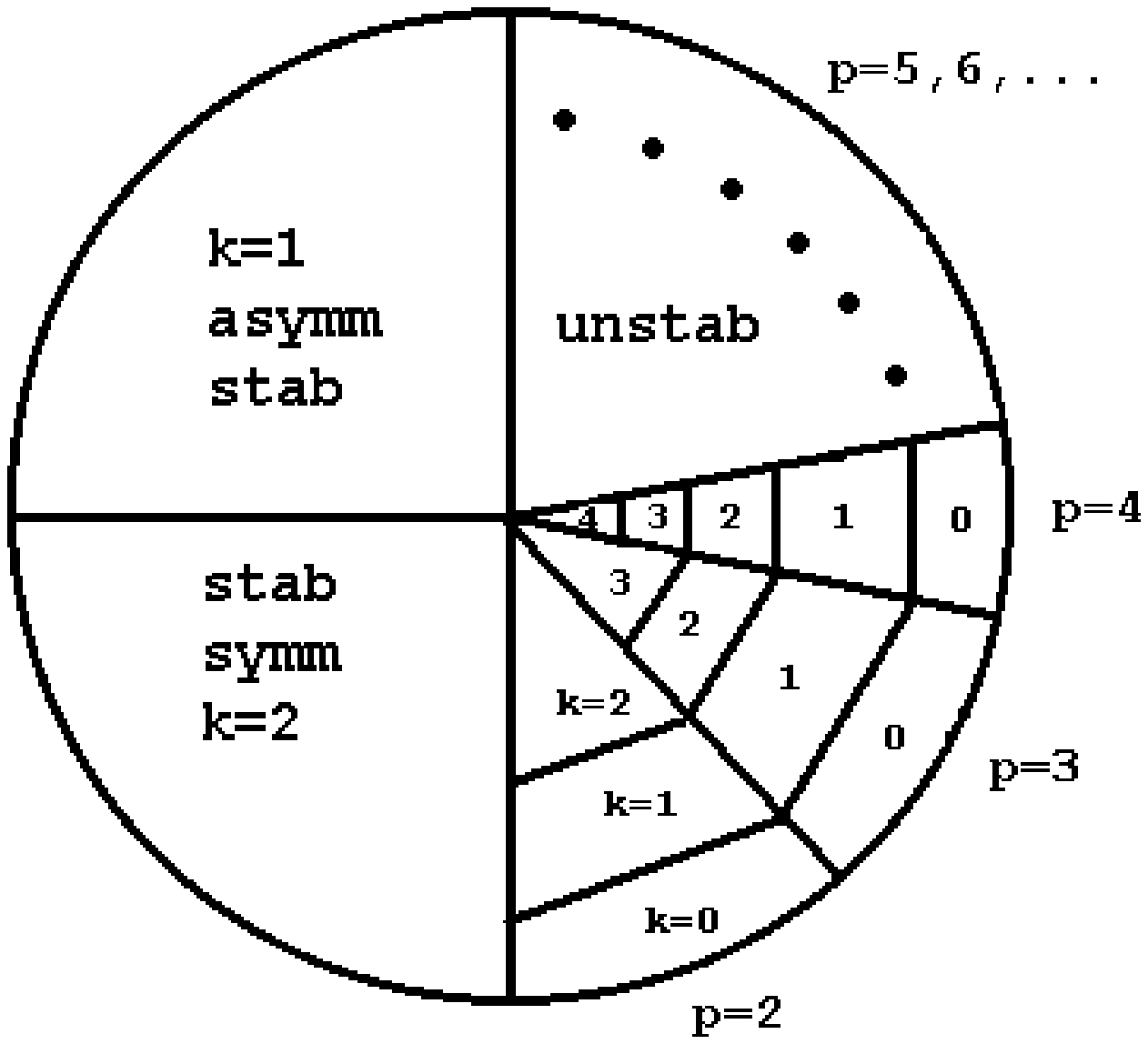}
\end{center}
\caption{The scheme of stratified structure of the
cycles.}\label{fig2}
\end{figure}

\begin{figure}
\begin{center}
\includegraphics*[scale=0.7]{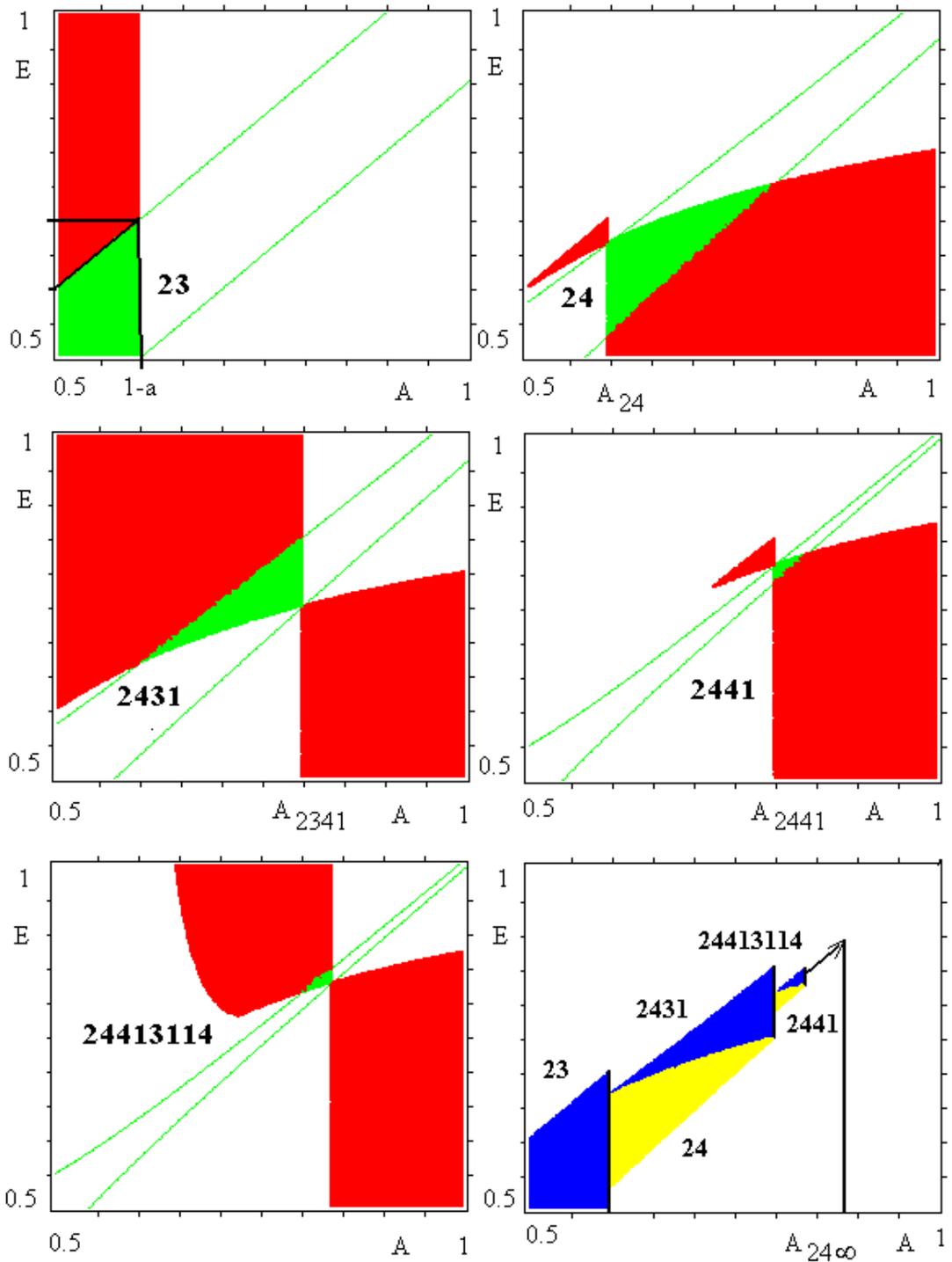}
\end{center}
\caption{The cycle diagram in the stability zone of the mapping.
On the plane of the control parameters (A, E) stable and unstable
cycles are mapped with the following coinciding routes: W=23;
W=24; W=2431; W=2441; W=24413114  and  the full 24-cascade of
doubling of the stable cycles experiencing "putting-out"
bifurcation with the change of the symmetry type.}\label{fig3}
\end{figure}

\begin{figure}
\begin{center}
\includegraphics*[scale=0.8]{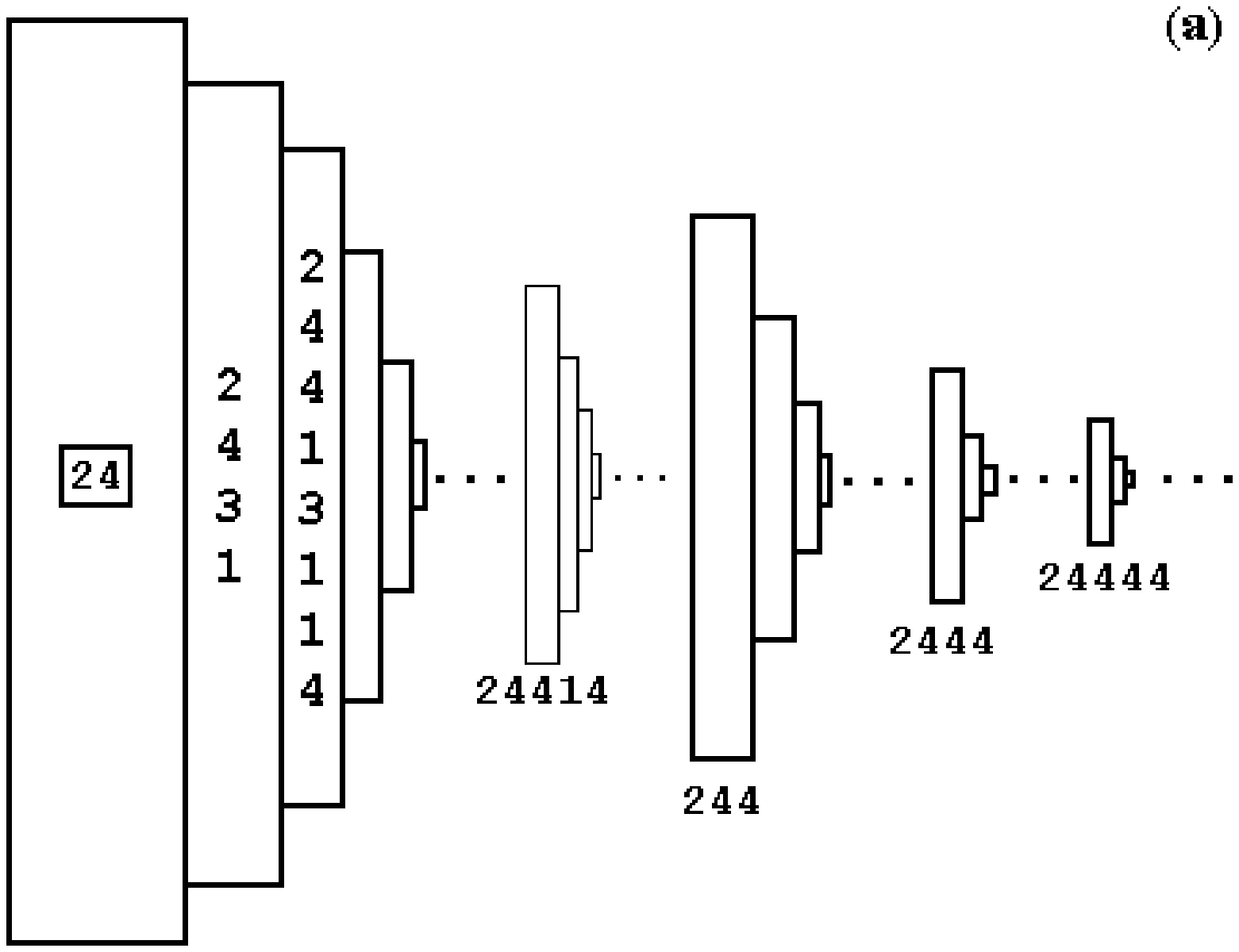}
\end{center}
\label{fig4a}
\end{figure}
\begin{figure}
\begin{center}
\includegraphics*[scale=0.6]{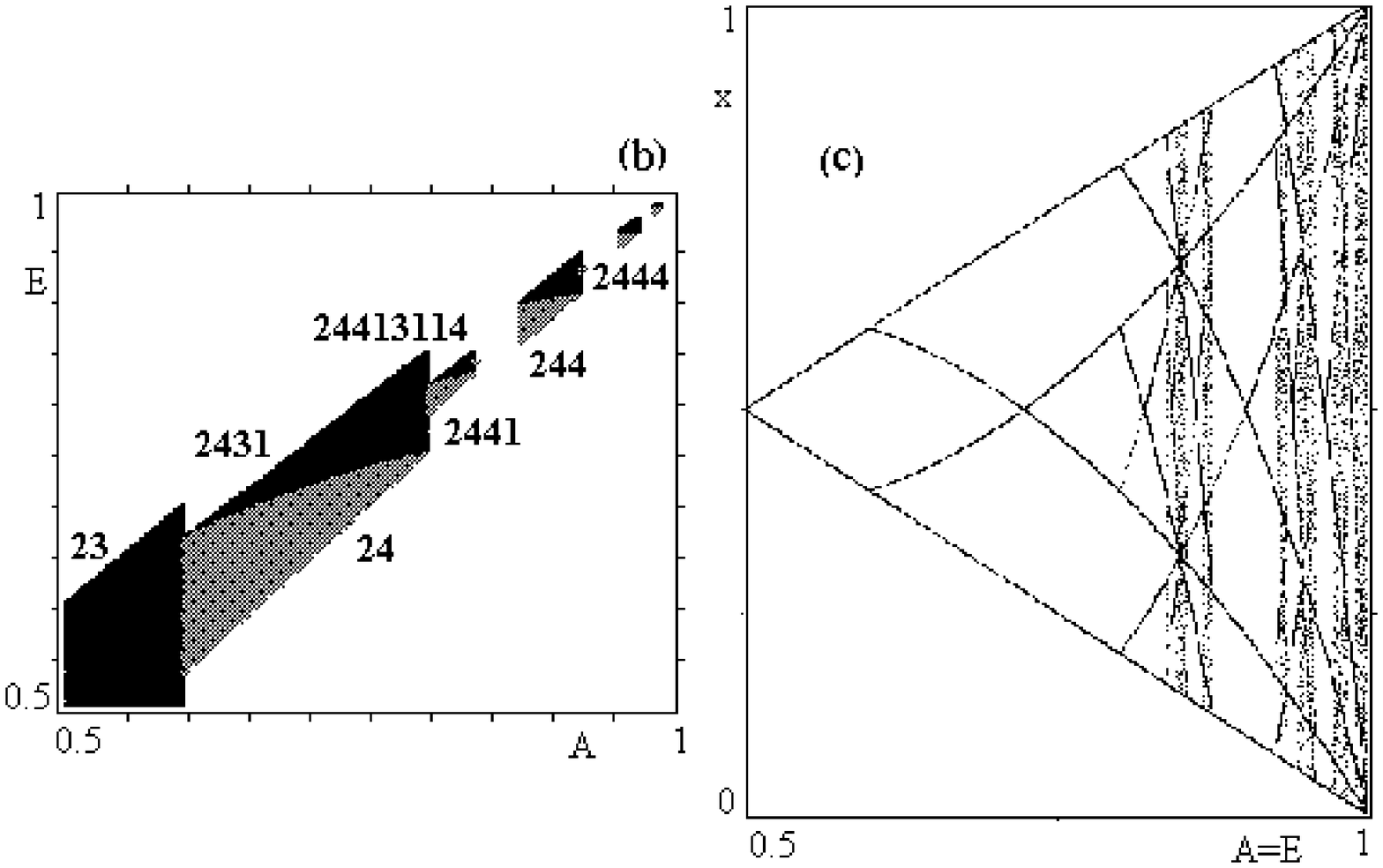}
\end{center}
\caption{The diagram of stable cycles: a) a telescopic structure
(schematically); b) the main (24; 244; 24…4; …) series of a
telescopic structure of the stable cycles; c) the skeleton of some
irreducible stable cycles, "generating" the cascades with period
doubling.}\label{fig4}
\end{figure}

\begin{figure}
\begin{center}
\includegraphics*[scale=1]{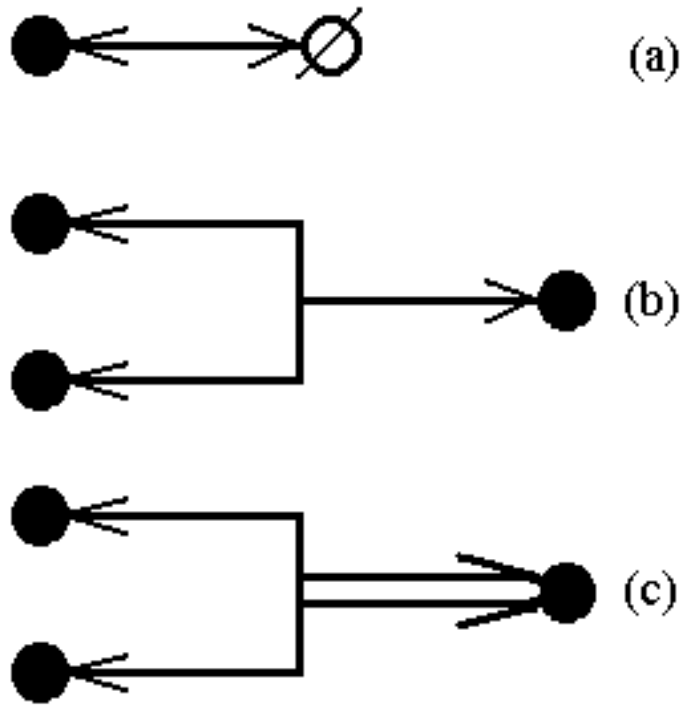}
\end{center}
\caption{ The cycles "putting-out" bifurcation types: a)
spontaneous "putting-out"; b) "puttings-out" without any changes
of period; c) "putting-out" with the period doubling.}\label{fig5}
\end{figure}

\end{document}